# Ultralong Efficient Photon Storage Using Optical Locking

Byoung S. Ham

Center for Photon Information Processing, and School of Electrical Engineering, Inha University,
253 Yonghyun-dong, Nam-gu, Incheon 402-751, S. Korea
bham@inha.ac.kr

Abstract: For the last decade quantum memories have been intensively studied for potential applications to quantum information and communications using atomic and ionic ensembles. With the importance of a multimode storage capability in quantum memories, on-demand control of reversible inhomogeneous broadening of an optical medium has been broadly investigated recently. However, the photon storage time in these researches is still too short to apply for long-distance quantum communications. In this paper, we demonstrate new physics of spin population decay dependent ultralong photon storage method, where spin population decay time is several orders of magnitude longer than the conventional constraint of spin phase decay time.

We demonstrate photon storage time far exceeding the critical constraint of spin phase decay time by adapting an optical locking method[1,2] to stimulated photon echoes[3], where photon echoes have intrinsic spatiotemporal multimode properties[1-12]. The dramatic extension of photon storage time has been achieved via population transfer-based coherence conversion between optical and spin states using the optical locking technique. Here we demonstrate photon storage nearly five orders of magnitude longer than that based on a rephasing halt such as in atomic frequency comb echoes[11]. Moreover, we demonstrate echo efficiency two orders of magnitude higher by adapting a phase conjugate scheme to solve the conventional echo reabsorption problem in a forward scheme[4]. The observed photon storage time can be extended up to several hours at cryogenic temperatures[13]. This breakthrough in photon storage time sheds light on long-distance quantum communications such as quantum cryptography relying on quantum repeaters[14].

Over the last decade, photon echoes[15] have been studied for multimode quantum memory applications owing to reversible inhomogeneous broadening in order to increase echo efficiency[4-6], remove spontaneous emission noise[7-10], and extend photon storage time[11]. Gradient[8-10], atomic frequency comb[7,11] (AFC), and controllable reversible inhomogeneous broadening techniques[4,6] have been introduced to increase photon echo efficiency or to remove spontaneous emission noise caused by excited atoms in two-pulse photon echoes[15]. To extend photon storage time, the stimulated (three-pulse) photon echo protocol has been widely used for classical all-optical information processing[3]. The stimulated photon echoes, however, lose the benefit of longer storage time for quantum memory applications due to population decay-dependent coherence loss. Unlike classical optical memories, quantum optical memory requires high-fidelity retrieval efficiency: Retrieval efficiency of 67% or more is needed for entanglement swapping used in quantum repeaters[16]. Recently, the optical locking method which stops the population decay-caused coherence loss in two-pulse photon echoes[4-6] has been adapted to a Raman scheme [see Ref. 17 for stimulated (on-resonance) Raman echoes], where storage time extends up to spin population decay time, several orders of magnitude longer than the optical counterpart[2].

Spontaneous emission noise in photon echoes caused by a rephasing optical pulse, however, creates a critical problem for single photon-based quantum memories. Although AFC and gradient echoes as modified photon echo protocols are free from the spontaneous emission noise, those techniques cannot be applied to long-distance quantum communications due to the short storage time confined by optical phase decay − much shorter than the spin phase decay constraint of one millisecond[7-11]. Even with the rephasing halt in AFC, the storage time extension is just one tenth of the optical phase decay time[11]. The spontaneous emission noise problem in conventional photon echoes for quantum memory applications, however, is resolved if squeezed light or multiphoton entangled light source is used due to propagation directionality (compared with omnidirectional spontaneous emission) and ultrashort window of echoes in an optical population decay time. Here, we demonstrate ultralong photon storage by adapting the optical locking method to conventional backward stimulated photon echoes to overcome the storage constraint in most quantum memory protocols and to solve the fundamental echo reabsorption problem in a forward scheme. The observed photon storage time far exceeds the critical constraint of spin phase decay time in quantum memory protocols such as slow-light based quantum memories[12,18-20], off-resonant Raman methods[21], and nitrogen vacancy diamonds[22,23].

The optical locking method originates in a spontaneous optical deshelving process applied to classical photon echoes to extend photon storage time[24]. Recently this idea of spontaneous atom deshelving has evolved[4,7] and has been developed for AFC echoes using coherent light[11]. The storage time extension in AFC echoes using identical π deshelving pulses is based on a rephasing halt[25], where the extended storage time is limited by spin inhomogeneous broadening, one tenth the optical phase decay time[11]. Thus, the goal of the present Letter is to present new physics of ultralong photon storage time to shed light on ultralong quantum memory applications for long-distance quantum communications, where at least one second storage time is required.

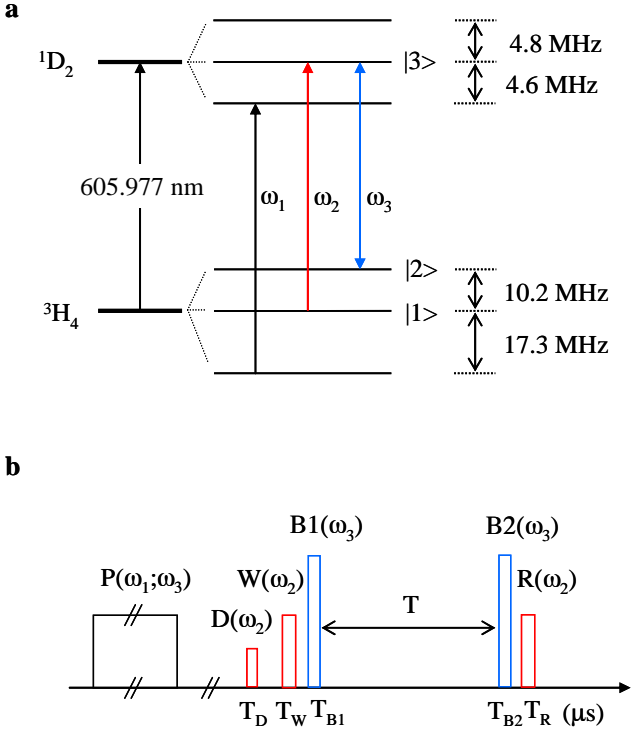

Figure 1. Schematic of optical locking. a, Partial energy level diagram of Pr:YSO at ~5 K. b, Light pulse sequence of a.

Figure 1a shows a partial energy level diagram of a rare earth $Pr^{3+}$ (0.05 at. %) doped $Y_2SiO_5$, and Fig. 1b represents the pulse sequence for optical locking in the stimulated photon echoes. The three laser beams in Fig. 1a are the modulated output of a ring-dye laser (Tecknoscan) pumped by a 532 nm laser (Coherent Verdi). Relative frequency adjustment for $\omega_1$, $\omega_2$, and $\omega_3$ is achieved by using acoustooptic modulators (Isomet) driven by radio frequency (rf) synthesizers (PTS 250). The light pulse P ($\omega_1;\omega_2$) is used for initial preparation of ground state population redistribution: $\rho_{11}=1$; $\rho_{22}=0$; $\rho_{33}=0$. The data pulse D in Fig. 2(a) and the inset of Fig. 3 is made by an arbitrary waveform generator (AWG610) for a hyperbolic secant pulse: $\Omega(t) = \Omega_0 \sech[\beta(t-t_0)]$, where $\Omega_0$ is the maximum Rabi frequency of the pulse and $\beta$ is real to determine the pulse duration. Other pulses are square pulses, so that a perfect population transfer by B1 and B2 cannot be made. Duration of each pulse of the laser beam is controlled by an rf switch (Minicircuits) and a digital delay generator (SRS DG 535). The repetition rate of the light pulse train is 10 Hz. The lights D, W, R, B1, and B2 are generated from the same laser beam. The lights D, W, and R copropagate as shown in Fig. 2. In Fig. 3, however, the light R oppositely propagates against W to form a phase conjugate scheme. The control lights B1 and B2 for optical locking are copropagating. The angle between D and B1 is 12.5 milliradians with overlap of ~90% (~80%) along the 1 mm (3 mm) sample. For initial atom preparation, both optical pulses at $\omega_1$ and $\omega_3$ are turned on for 10 ms, so that ground state atoms are incoherently pumped onto state |1> ($\pm3/2$, $^3H_4$). Optical locking is achieved by the deshelving optical pulse pair B1 and B2 at $\omega_3$. To satisfy the phase recovery condition in an optically dense medium, B1 and B2 must satisfy the following conditions[1,25,26]:

$$\Phi_{B1}+\Phi_{B2}=4n\pi, \quad (1\text{-}1)$$
$$\Phi_{B1}=(2n-1)\pi, \quad (1\text{-}2)$$

where $\Phi$ is the pulse area, and n is an integer. In Fig. 1b, the data photon D is stored in a spectral grating generated by the subsequent W, where both ground and excited atoms are phase locked according to the stimulated photon echo theory[3] (see also Supplementary Fig. S1). Here, the spectral grating is a population modulation in the optical inhomogeneous broadening. All spectrally grated atoms in the ground (|1>) and excited (|3>) states are equally distributed and form a frequency comb, but in an exactly opposite manner at detuning $\delta$ from line center:

$$\rho_{33}^\delta = 1 - \rho_{11}^\delta; \quad \sum_{\delta=-\infty}^{\infty} \rho_{11}^\delta = \sum_{\delta=-\infty}^{\infty} \rho_{33}^\delta, \quad (2)$$

where $\rho_{ii}$ is the normalized population of state |i>.

For the optical locking experiment, we compare a forward propagation scheme in Fig. 2 with a backward propagation scheme in Fig. 3. The purpose of the backward propagation scheme[4,6] is to let echoes retrace backward along the data trajectory to solve the echo reabsoprtion problem in a forward scheme[4,27]. Since the function of B1 and B2 is for optical deshelving of the phase locked atoms in state |3>, the phase matching condition is the same as for the conventional stimulated photon echoes[3]:

$$\vec{k}_E = -\vec{k}_D + \vec{k}_W + \vec{k}_R, \quad (3)$$
$$\varpi_E = -\varpi_D + \varpi_W + \varpi_R, \quad (4)$$

where $\vec{k}_i$ and $\varpi_i$ are wave vector and angular frequency of the pulse $i$, respectively. Unlike the optical locking applied to the rephasing process[1,11,25], the optical locking applied to the spectral grating in Fig. 1 freezes the phase decay process of individual atoms from both optical and spin dephasing. Thus, unlike Ref. 11, the spin dephasing-independent photon storage time should be expected, where ultralong storage time extension via a perfectly collective coherence swapping between optical and spin transitions is possible even in a spin jitter-based noisy optical system (via spin inhomogeneous broadening). Note here that the photon echo methods[1-12] have been accepted as quantum memory protocols because, in a weak-field limit, a classical data pulse can be treated quantum mechanically[28].

Figure 2 shows experimental demonstrations of the optically locked echoes in a forward propagation scheme. The optical locking pulses B1 and B2 copropagate at a slight angle against the other group of copropagating pulses of D, W, and R: $k_E = k_D$. Figure 2a shows a plot of conventional three-pulse photon echoes as a function of ΔT (R delay from W) without B1 and B2 for a reference. The measured decay time τ is temperature independent, where τ represents the optical population decay time $T_1^{Opt}$ ($T_1^{Opt}$=160 μs; see Ref. 29) according to the stimulated photon echo theory[3]. The inset shows one of the data points used for the plot, where the third peak is the leaked two-pulse photon echo signal, and the last peak is the stimulated photon echo. The two-pulse photon echo signal comes either from coherence leakage or imperfect pulse area geometry due to a Gaussian profile of the light in a cross section. The measured maximum echo efficiency (echo intensity ratio to the data intensity at ΔT=0) is about 0.2 % due to the well known echo reabsorption.

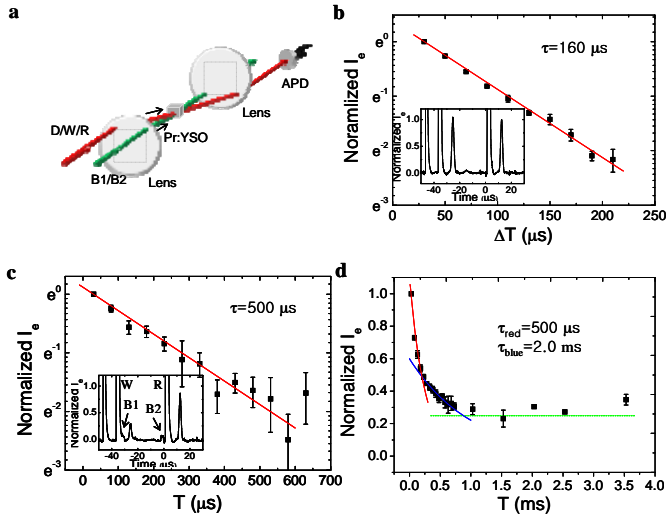

Figure 2. Observations of ultralong optically locked echoes in a forward propagation scheme. a, Stimulated photon echoes without B1 and B2 as a function of delay ΔT (R delay from D) in Fig. 1. b and c, Optically locked echoes with B1 and B2 as a function of delay T in Fig. 1. The delays of R from B2 and W from D are fixed at 2 μs. Red lines are the best fit curves for exp(−2t/τ). The pulse duration of D, W, and R in Fig. 1 is set at 1 μs for the π/2 pulse area and 4.6 mW in power. The pulse duration of control pulse B1 (B2) is set at 1.2 μs (3.6 μs) for the π (3π) pulse area at 14.5 mW in power. The optimum power of the light B1 and B2 is predetermined by Rabi flopping measurement. An avalanche photodiode APD1 detects light in the line of D and R including conventional photon echoes, and the detected signals are directly fed into a digital oscilloscope and recorded by averaging 10 samples. The spot diameter [exp(−2) in intensity] of B1, B2, D, W, and R are ~300 μm. The Pr:YSO sample in a liquid helium cryostat (Advanced Research System) is kept at a temperature of ~5 K. The resultant absorption of a data pulse D is ~70%, where optical depth d (d=αl, where l=1 mm) is 1.0. All light pulses are vertically polarized, and propagate along the crystal axis of the medium (Pr:YSO). The angle between beams D and B1 is 12.5 miliradians and overlapped by 90% through the sample ($Pr^{3+}$:$Y_2SiO_5$) of 1 mm in length. The dashed green line indicates 50% echo efficiency in amplitude. The measured optical phase decay time ($T_2$) is 25 μs at ~5 K.

Figure 2b shows optically locked echoes (last peak in the inset) as a function of delay time T (B2 delay from B1) with the optical locking pulses B1 and B2, where the two-pulse photon echo (third peak) is severely suppressed due to population transfer to spin state |2>. Scattering of B1 and B2 is to denote their relative positions. The measured echo decay time τ in a short time scale, T < $T_2^{Spin}$ (spin phase decay time $T_2^{Spin}$ is 500 μs; see Fig. 4 of Ref. 17) is τ=500 μs, nearly two orders of magnitude longer than that in AFC echoes based on a rephasing halt[11]. Unlike AFC echoes limited by spin inhomogeneous broadening, this increment in photon storage time proves that the optical locking applied to the spectral grating of the stimulated photon echoes freezes individual atom phase evolution resulting in spin inhomogeneous broadening independence. However, due to imperfect population transfer in the present dilute sample (measured optical depth d is d=αl=1.0) as mentioned in Ref. 25, the remnant atoms on state |3> contribute to coherence leakage showing faster decay time resulting in a transient phenomenon toward a long time scale (will be discussed in Fig. 3): See Supplementary Fig. S2.

In a long time scale (see Fig. 2c), the decay process of the optically locked echo signals gradually slows and eventually saturates at a certain level: The green dashed line indicates 25% of the maximum echo intensity. In an intermediate region, 300 μs < T < 700 μs, the measured decay time is τ=2 ms (blue curve), which is far beyond $T_2^{Spin}$. The flat line of saturation is due to new physics of photon storage based on spin population decay time $T_1^{Spin}$ ($T_1^{Spin}$ is 100 sec at T=1.6 K, but exponentially decreases as temperature increases; see Ref. 29). In the transient regime (0<T<1 ms), the varying decay phenomenon can be explained as T dependent remnant atoms causing coherence loss[26]. Compared with rephasing control using optical locking[11,25], the photon storage time extension in Fig. 2c is several orders of magnitude longer.

Figure 3 shows experimental demonstrations of the optically locked echoes in a backward (phase conjugation) scheme (see the inset): $k_R = -k_W$; $k_E = -k_D$. To support the remnant population caused coherence loss discussed in Fig. 2b, we use a longer sample (l=3 mm) to increase optical depth d (d=αl=2.4: see Supplementary Fig. S2), where the population transfer rate increases. According to the noncollinear propagation scheme, the laser beams' spatial overlap inside the sample is ~80%, which should cause reduced echo efficiency. Due to the increased optical depth, the stimulated echoes without B1 and B2 are severely reduced by the echo reabsorption process[27] (not shown): If B1 and B2 are switched on, the stimulated echoes completely disappear. Figures 3a and 3b show experimental data. The first data (red dotted circle) at t=0 in Fig. 3b is due to imperfect population transfer. This phenomenon, however, should disappear if T>$T_1^{Opt}$ because of no remnant population remains on the excited state. The measured echo efficiency (maximum echo intensity ratio to the data intensity) in Fig. 3 is 10%, nearly two orders of magnitude higher than that of Fig. 2. This enhancement factor of 50 is the first experimental proof of the backward photon echo scheme proposed in Ref. 4: see also Supplementary Fig. S3. Compared with Fig. 2, the transient phenomenon in Fig. 3 gets much slowed down. The measured echo decay time $τ_1$ in a short time scale (T<$T_2^{Spin}$) is $τ_1$= 2.8 ms, which is much longer ($\tau$= 0.5 ms) than that of Fig. 2. Hence, the intermediate transient phenomenon can be explained as remnant population induced coherence loss in an optically dilute sample. In an optically dense medium, however, the transient phenomenon would disappear.

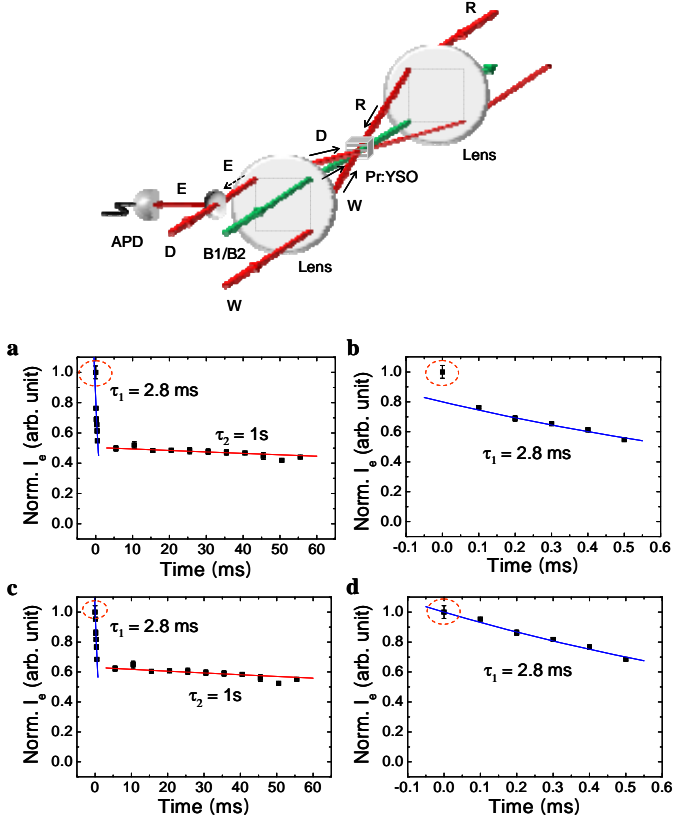

Figure 3. Observations of ultralong optically locked echoes in a backward propagation scheme . Inset: backward propagation scheme. a and b, Optically locked echoes with B1 and B2 as a function of delay T in Fig. 1. The delays of R from B2 and W from D are fixed at 2 μs. c and d, Adjusted data of a and b, respectively. The first data is adjusted to 80%, and all other data are normalized to the one. Red lines are the best fit curves for exp(−2t/$\tau$). The angle between beams D and B1 is 12.5 milliradians and overlapped by 80% through the sample ($Pr^{3+}$:$Y_2SiO_5$) of 3 mm in length. The resultant absorption of a data pulse D is 90%, where optical depth d (d=$\alpha l$, where $l$=1 mm) is 2.4. The delay W from D is 5 μs. Each spot diameter [exp(−2) in intensity] of B1 and B2 is 340 μm. The spot diameter [exp(−2) in intensity] of D, W, and R is 280, 390, and 590 μm, respectively. The peak power of D, W, R, B1 and B2 is 7, 23, 10, 12, 12 mW. The pulse duration of D, W, R, B1, and B2 is 0.76, 0.7, 2, 1.2, and 3.6 μs, respectively. The Pr:YSO sample in a liquid helium cryostat (Advanced Research System) is kept at a temperature of ~6 K.

In a long time scale (T>>$T_2^S$), the measured echo decay time $\tau_2$ ($\tau_2$=1 s) is ultralong, where $\tau_2$ represents the spin population decay time $T_1^{Spin}$ at 6K (Ref. 29). Another interesting result is that the photon storage time, taken until the echo intensity drops to 50% of its maximum value, is near 100 ms. Here, the 50% echo efficiency indicates a no-cloning regime without post-selection[30], where quantum error correction protocols become available[9]. Considering coarse experimental conditions with 300 kHz laser jitter and imperfect population transfer in a dilute sample, the observation in Fig. 3 attests a breakthrough in the spin phase limited photon storage time for potential applications of ultralong quantum memories, where new photon storage time is governed by spin population decay time. The use of an optically dense medium is important to achieving higher fidelity because of full absorption of data photons, near perfect retrieval efficiency in a phase conjugation scheme, and complete population transfer by the optical locking pulses to remove the coherence loss-based transient phenomenon.

For a very weak data pulse or single photons, most atoms stay in the ground state after interaction with the data pulse D. By the rephasing pulse in a two-pulse photon echo scheme, or by the read pulse R in a three-pulse photon echo scheme, the system becomes inverted resulting in possible spontaneous emission noise. In spite of both a shorter echo window than the spontaneous emission decay time, and pencil-like echo directionality, even a few-photon level of spontaneous emission noise can severely decrease fidelity. However, the use of squeezed light or multi photon entangled light should practically remove the spontaneous emission noise problem.

In conclusion, we proposed and experimentally demonstrated an ultralong photon storage protocol using optical locking applied to stimulated photon echoes, where the storage time far exceeds the critical constraint of spin phase decay time in conventional quantum memories. The observed photon storage extension time (1 sec) using optical locking was five orders of magnitude longer than that ($10^{-5}$ sec) of AFC echoes based on rephasing halt. Moreover, the observed higher echo efficiency (x50) in a backward propagation scheme was a direct proof to solve the echo reabsorption problem in a conventional forward propagation scheme. The population inversion-caused spontaneous emission noise can be circumvented if squeezed light or multiphoton entangled light is used as a data source. Thus, the present method holds potential for quantum repeaters in long-distance quantum communications owing to multimode ultralong photon storage.

This work was supported by the CRI program (Center for Photon Information Processing) of the Korean government (MEST) via National Research Foundation, S. Korea. BSH acknowledges that J. Hahn contributed to the experiments.